\documentclass[12pt]{iopart}
\usepackage{iopams}  

\usepackage{graphicx}
\usepackage{dcolumn}% Align table columns on decimal point
\usepackage{bm}% bold math

\usepackage{color}
\usepackage{morefloats}

\newcommand{\ep}{\varepsilon}

\newcommand{\equa}{{equation }}

\begin{document}

\title[Relative entropy and distribution functions of P{\'o}lya Urn processes]{Understanding frequency distributions of 
path-dependent processes with non-multinomial maximum entropy approaches
}

\author{Rudolf Hanel$^1$, Bernat Corominas-Murtra$^1$, and Stefan Thurner$^{1,2,3,4}$}

\address{$^1$Section for Science of Complex Systems, Medical University of Vienna, Spitalgasse 23, 1090 Vienna, Austria}
\address{$^2$Santa Fe Institute, 1399 Hyde Park Road, Santa Fe, NM 87501, USA}
\address{$^3$IIASA, Schlossplatz 1, A-2361 Laxenburg, Austria}
\address{$^4$Complexity Science Hub Vienna, Josfst{\"a}tterstrasse 39, A-1090 Vienna, Austria}
\ead{stefan.thurner@meduniwien.ac.at}
\vspace{10pt}
\begin{indented}
\item[]\today
\end{indented}

\begin{abstract}
Path-dependent stochastic processes are often non-ergodic and observables can no longer be computed 
within the ensemble picture. The resulting mathematical difficulties pose severe limits to the 
analytical understanding of path-dependent processes. Their statistics is typically 
non-multinomial in the sense that the multiplicities of the occurrence of states is not a multinomial
factor. The maximum entropy principle is tightly related to multinomial processes, 
non-interacting systems, and to the ensemble picture; It loses its meaning for path-dependent processes.
Here we show that an equivalent to the ensemble picture exists for path-dependent processes,  
such that the non-multinomial statistics of the underlying dynamical process, by construction, 
is captured correctly in a functional that plays the role of a 
% RUDI in
relative 
% RUDI out
entropy. We demonstrate this for 
self-reinforcing Polya urn processes, 
% RUDI in
%that explicitly break
% multinomial structure. 
which explicitly generalise
% RUDI out
multinomial statistics. 
We demonstrate the 
% RUDI in
%new method 
adequacy of this %novel 
constructive approach towards non-multinomial 
pendants of entropy
% RUDI out
by computing frequency and rank distributions of Polya urn processes. 
% RUDI in
%For the first time we are able to use detailed 
We show how  
% RUDI out
microscopic update rules of a path-dependent process 
% RUDI in
%to 
allow us to explicitly
% RUDI out
construct a non-multinomial entropy 
functional, that, when maximized, predicts the time-dependent distribution function. 
\end{abstract}

%\pacs{02.70.Rr, 02.50.Cw, 05.10.Gg}

%\submitto{\NJP}
%\maketitle

\section{Introduction}
``It seems questionable whether the Boltzmann principle alone, meaning without a complete [...] mechanical description or some other complementary 
description of the process, can be given any meaning.'' 
Einstein's famous critical comment on the completeness of Boltzmann entropy,  \cite{Einstein:1910}, is still thought provoking.
For ergodic systems, e.g. \cite{LandauLifshitz}, over a well defined set of states, 
this critique has turned out to be of minor relevance. 
Here we demonstrate how Einstein's observation becomes relevant
again when dealing with non-ergodic, path-dependent systems or processes,
i.e. processes where ensemble and time averages cease to yield identical results
and the ensemble descriptions of a processes fails to describe the dynamics 
of a particular process (e.g. compare \cite{OPeters}).

% Rudi in
Moreover, for path dependent systems we have to specify 
what we mean with ``entropy'', since no unique generalization of entropy 
from equilibrium to non-equilibrium systems exists.
However, Boltzmann's principle is grounded in the idea
that in large systems the most likely samples we may draw from a process, 
i.e. the so called {\em maximum-configuration},
also characterize the typical samples, while it becomes very unlikely to 
draw atypical samples.
In fact we will demonstrate the possibility to
directly construct ``entropic functionals'' from the microscopic properties determining
the dynamics of a large class of non-ergodic processes using maximum-configuration frame work.
In this approach we identify relative entropy (up to a multiplicative constant) 
with the logarithm of the probability to observe a particular macro state 
(which typically is represented by a histogram over a set of observables states), 
compare e.g. \cite{HTMGM3}. 
% Rudi out
By construction, maximisation of the resulting entropy
functionals leads to adequate predictions of statistical properties of non-ergodic processes, in maximum configuration.

For ergodic processes it is possible to replace time-averages of observables by their ensemble-averages, 
which leads to a tremendous simplification of computations.  
%In particular, the use of entropy and its maximization under constraints has become
%a standard procedure for understanding and predicting distribution functions of large systems in equilibrium. 
%In such applications the notion of {\em ergodicity} is crucial, meaning that observed probabilities essentially coincide with prior probabilities. 
%This is certainly 
In particular, this is true for systems composed of independent particles or for Bernoulli processes, i.e. processes where samples are drawn independently, 
and the states of the independent components or observations collectively follow a multinomial statistics. The multinomial statistics of 
such a system with $W$ observable states
$i=1,\cdots,W$ is captured by 
% Rudi in
a functional that coincides with
% Rudi out
Shannon entropy \cite{Shannon1948}, 
$H(p)=-\sum_{i=1}^W p_i \log p_i$. In this context $p=(p_1,\cdots,p_W)$ is the empirical {\em relative frequency}
distribution of observing states $i$ in an experiment of drawing from the process for $N$ times, i.e. $p=k/N$ is the normalized {\em histogram}
of the experiment where state $i$ has been drawn $k_i$ times.
%count the number of times $i$ gets sampled after $N$ observations.
% Rudi in
Clearly, $\sum_i k_i=N$.
% Rudi out
In this context $H(p)$ can be understood as the logarithm of the multinomial factor, 
i.e. $-\sum_{i=1}^W p_i\log p_i\sim \frac1N \log {N\choose k}$, where 
%$N$ is the number of iterations, 
and ${N\choose{k}}=N!/\prod_{i=1}^W k_i!$ (e.g. compare \cite{Jaynes1968}). 

Maximization of Shannon entropy under constraints therefore is a way of finding the most likely 
% Rudi in
relative frequency
% Rudi out
distribution function (normalized histogram of sampled events) one will observe when measuring a system, 
provided that it follows a multinomial statistics. 
Constraints represent knowledge about the system. 
% Rudi in
%as moments of the distribution function. 
Bernoulli processes with multinomial statistics are characterized by the prior probabilities, $q=(q_1,\cdots,q_W)$.
%, of a sampling 
%observable states $i=1,\cdots,W$.
% Rudi out
In general, the set of parameters characterizing a process, we denote by $\theta$.
In the multinomial case $\theta\equiv q$.
%Suppose we perform an experiment and measure every state $i$, say $k_i$ times. The total number of measurements we denote 
%$N=\sum_i k_i$.  We call $k=(k_1,\cdots,k_W)$ the {\em histogram} of the experiment, $p=k/N= (p_1,\cdots,p_W)$  
%is the empirical {\em relative frequency} distribution.

Denoting the probability to measure a specific histogram by $P(k|\theta,N)$, 
the most likely histogram $\hat k$, that maximizes $P(k|\theta,N)$, is the optimal predictor or the so-called
{\em maximum configuration}.
For a multinomial distribution function, $P(k|\theta,N)={N\choose{k}}q_i^{k_i}$, where  $q_i$ are the prior probabilities (or biases), 
the functional that is maximized is $\psi(p|\theta)=H(p)+\sum_ip_i\log q_i$, which is (up to a sign) 
called the {\em relative entropy} or Kullback-Leibler divergence \cite{KullbackLeibler}. 
The term $H(p)$ coincides with Shannon entropy, the term that depends on $q$ is called {\em cross-entropy} and is a linear functional in $p$.
By re-parametrizing $q_i=\exp(-\beta\ep_i)$, where $\beta>0$ is a constant, 
one gets the standard max-ent functional
\begin{equation}
	\psi(p|\theta,N)=H(p)-\beta\sum_ip_i\ep_i \qquad \left[\sum_i p_i=1 \right] \quad . 
\label{standardMEP}
\end{equation}
In statistical physics, the constants $\ep_i$ typically correspond to energies and $\beta$ to the so called {\em inverse temperature} of a system.
Maximization of this functional with respect to $p$ yields the most likely empirical distribution function;  
this is sometimes called the {\em maximum entropy principle} (MEP).  

\begin{figure}[t]
	\centering
		\includegraphics[width=0.7\columnwidth]{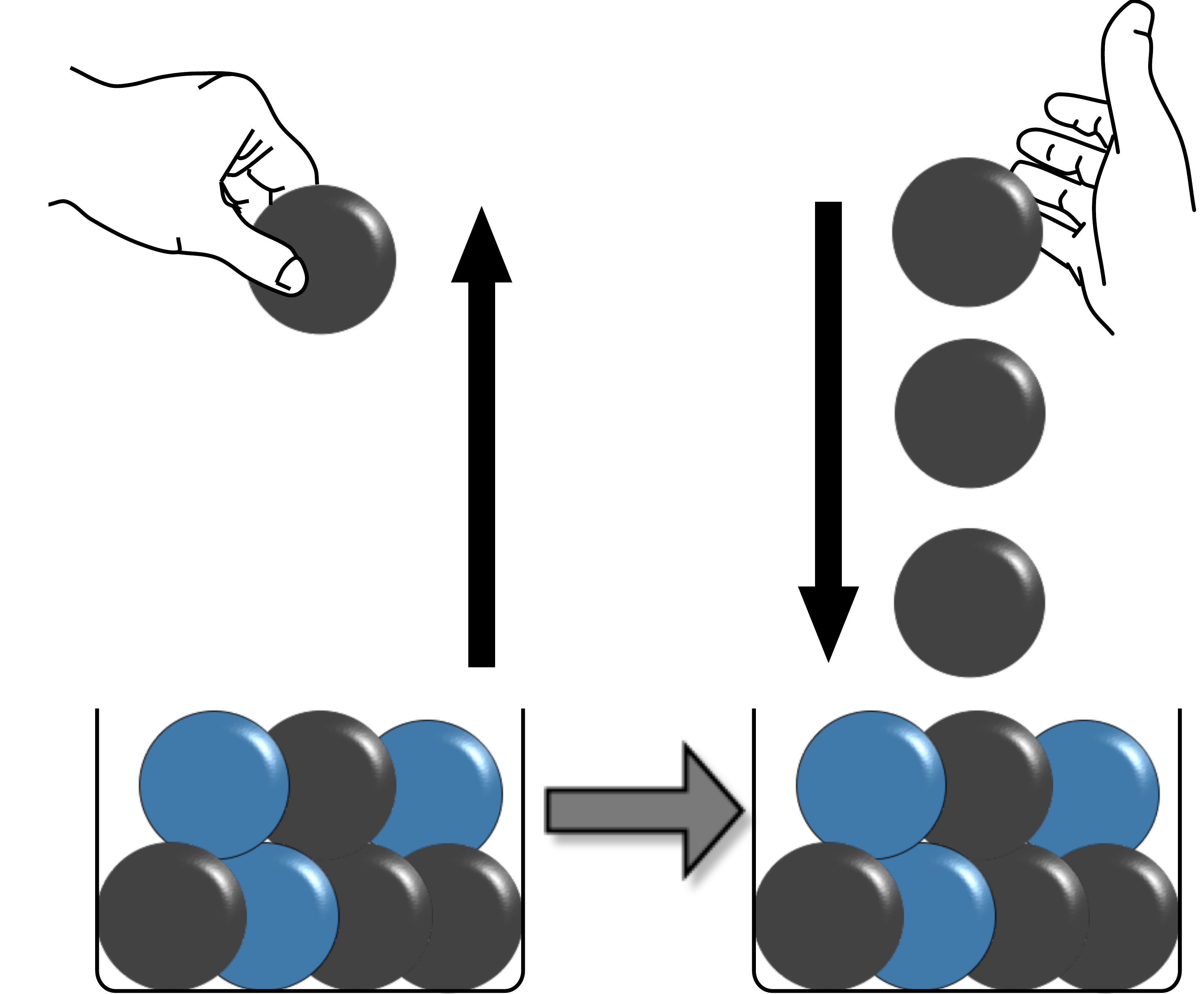}
	\caption{Schematic illustration of a P{\'o}lya process. When a ball of a certain color is drawn, 
	it is replaced by $1+\delta$ balls of the same color. Then the next ball is drawn and the process 
	is repeated for $N$ iterations. Here $\delta =2$. 
	This reinforcement process creates a history-dependent dynamics. 
	The configurations obtained after successive iterations have non-multinomial structure.
	\label{fig:PolyaUrn}
	}
\end{figure}

Clearly, systems composed of independent components follow a multinomial statistics. %, $P(k|\theta,N)$.  
Note that a multinomial statistics is also a direct consequence of working with ensembles of
statistically independent systems. In this case the multinomial distribution function reflects the ensemble 
property and is not necessarily a property of the system itself.  
Therefore $H(p)$ only has physical relevance for systems that consist of sufficiently independent elements.
For path-dependent processes, where ensemble- and time-averages typically yield different results, $H(p)$ remains the
entropy of the ensemble picture, but ceases to be the ``physical'' entropy that captures the time evolution of a path-dependent process. 
Obviously, assuming that the entropy functional $H$, which is consistent with an underlying multinomial statistics,
in general also is adequate for characterizing path-dependent processes that are inherently 
non-multinomial (break multinomial symmetry), is nonsensical.   

Surprisingly, the possibility that non-multinomial max-ent functionals can be constructed for path-dependent processes 
%-- as an alternative to the ensemble picture frequently invoked for ergodic systems--  
seems to have 
% Rudi in
%been overlooked. 
caught only little attention.
% Rudi out
In \cite{HTMGM3} it was noticed that a particular class 
of non-Markovian random walks with strongly correlated increments can be constructed, 
where the multiplicity of event sequences is no longer 
given by the multinomial factor,
%multinomial,  
and the max-ent entropy functional of the process class exactly violates the composition 
axiom of Khinchin \cite{Khinchin}. The general method of constructing a relative entropy principle 
for a particular process class does not inherently depend on the validity of particular 
information theoretic axioms, 
which opens a way for a general treatment of path-dependent, and non-equilibrium processes. 
We demonstrate this by constructing the max-ent entropy of
multi-state {\em P{\'o}lya urn processes}, 
\cite{Polya1923, Polya:1930} 
%We demonstrate how a max-ent approach can be successfully constructed for a wide class of non-multinomial processes. 

In multi-state P{\'o}lya processes, once a ball of a given color is drawn from an urn, it is replaced by a number of $\delta$ 
balls with the same color. They represent self-reinforcing, path-dependent processes that display 
the {\em the rich get richer} and {\em the winner takes all} phenomenon. 
P{\'o}lya urns are related to the beta-binomial distribution, Dirichlet processes, the Chinese restaurant problem, 
and models of population genetics. Their mathematical properties were studied in \cite{Wallstrom2012, Johnson:1977}, 
extensions and generalizations of the concept are found in \cite{Mahmoud:2008,Kotz2000}, applications to  
limit theorems in \cite{Janson2004,Smythe1996,Gouet1993}. 
P\'olya urns have been used in a wide range of practical applications including response-adaptive clinical trials \cite{Tolusso2011}, 
tissue growth models \cite{Binder2009}, institutional development \cite{Crouch2004}, computer data structures \cite{Bagchi1985}, 
resistance to reform in EU politics \cite{Geppert2012}, aging of alleles and Ewens's sampling formula \cite{Donnelly1986,Hoppe1984}, 
image segmentation and labeling \cite{Banerjee1999}, and the emergence of novelties in evolutionary scenarios \cite{Alexander:2012, Tria:2014}.
A notion of P\'olya-divergence was recently defined in \cite{Grendar2010} in the context of Sanov's theorem \cite{Sanov1957},
% RUDI in
which may be regarded as a notion of P\'olya-divergence for very small reinforcement parameter $\delta$. 
We use a constructive approach towards identifying the maximum configuration of 
sampled histograms, i.e. towards identifying the most likely distribution function, one may observe. 
%Our constructive approach proceeds along very different lines, yielding a functionally distinct notion of divergence, i.e. relative entropy.
Our approach allows us to access strong reinforcement and the transition of P{\'o}lya urn dynamics from
Bernoulli-process like behavior to a winner-takes-all type of dynamics can be studied. 
% RUDI out

%%%%%%%%%%%%%%%%%%%%%%%%%%%%%%%%%%%%%%%%%%%%%%%%%%%%%
\section{Non-multinomial max-ent functionals}
%%%%%%%%%%%%%%%%%%%%%%%%%%%%%%%%%%%%%%%%%%%%%%%%%%%%%

% Rudi in
The general aim is to construct a max-ent functional for a path-dependent process,
which allows us to infer the maximum configuration, i.e. the most likely sample we may draw
from a process of interest.
% Rudi out
From a given class of processes $X$ we select a particular process $X(\theta)$, specified by
a set of parameters, $\theta$. Running the processes $X(\theta)$ for $N$ consecutive iterations produces a 
sequence of {\em observed states} $x(\theta,N)=[x_1,\cdots,x_N]$, 
where each $x_n$ takes a value from $W$ possible states. 
As before, we assume the existence of a most likely histogram $\hat k$, that maximises $P(k|\theta,N)$.
To construct a max-ent functional for $X$, one has to conveniently rescale $P(k|\theta,N)$, which 
happens in two steps.  First, we define $\Psi(p|\theta,N)\equiv\log P(N p|\theta,N)$.
% Rudi in
Note, if $\hat k$ maximises $P(k|\theta,N)$, then
$\hat p=\hat k/N$ maximises $\Psi(p|\theta,N)$. 
One may interpret $-\Psi$ as a measure of probability in bits.
% Rudi out
Second, a scaling factor $\phi(N)$ can be used to scale out the leading term of the $N$ dependence of $\Psi$. 
Typically $\phi(N)=N^c$, for some constant $1\geq c>0$, compare \cite{HTMGM3}. $\phi(N)$ and corresponds to the
{\em effective} number of degrees of freedom of samples of size $N$.
We {\em identify} the max-ent functional with $\psi(p|\theta,N)\equiv\Psi(p|\theta,N)/\phi(N)$.
Again, if $\hat k$ maximises $P(k|\theta,N)$ with $\sum_i k_i=N$, then $\hat p=\hat k/N$ maximises $\psi(p|\theta,N)$, with $\sum_i p_i=1$. 
% RUDI in
In other words, $\psi(p|\theta,N)$ represents (up to a sign) a functional providing us with a notion of 
{\em relative entropy} (information divergence) for the process-class $X$.
If this process-class $X$ is the class of Bernoulli-processes, such that $P(k|q,N)$ is the multinomial distribution, 
then asymptotically $-\psi(p|q,N)\sim \sum_i p_i( \log p_i - \log q_i )$, is the Kullback-Leibler divergence, and $\phi(N)=N$.
In the following we compute $\psi(p|\theta,N)$ for P{\'o}lya urn processes.
% RUDI out 

%%%%%%%%%%%%%%%%%%%%%%%%%%%%%%%%%%%%%%%%%%%%%%%%%%%%%
\section{Max-ent functional for P\'olya urns}
%%%%%%%%%%%%%%%%%%%%%%%%%%%%%%%%%%%%%%%%%%%%%%%%%%%%%

In urn models observable states $i$ are represented by the colors balls contained in the urn can have.
The likelihood of drawing a ball of color $i$ is determined by the number of 
balls contained in the urn. Initially the urn 
contains $a_i$ balls of color $i=1,\cdots,W$. The {\em initial prior probability} to 
draw a ball of color $i$ is given by $q_i=a_i/A_0$, where $A_0=\sum_i a_i$ is the total 
number of balls initially  in the urn.
Balls are drawn sequentially from the urn. Whenever a ball of color $i$ is drawn, it is put back into the urn 
and another $\delta$ balls of the same color are added. This defines the multi-state P\'olya process \cite{Polya1923}.
A particular P\'olya process is fully characterised by the parameters, $\theta=(q_1,\cdots,q_W;A_0,\delta)$. 
Drawing without replacement is the {\em hypergeometric} process, drawing with replacement ($\delta=0$), is the {\em multinomial} process.  

If $\delta>0$, after $N$ trials there are $a_i(N)=a_i+\delta k_i$ balls of color $i$ in the urn ($a_i=a_i(0)$). The 
total number of balls is $A(N)=\sum a_i(N)=A_0+N\delta$, 
and the probability to draw a ball of color $i$ in the $(N+1)$th step is 
\begin{equation}
	p(i|k,\theta)=\frac{a_i(N)}{A(N)}=\frac{a_i + k_i\delta}{A_0+N\delta}\,,
\end{equation}
which depends on the history of the process in terms of the histogram $k$. With 
$x(0)=[\,\,]$ the {\em empty sequence}, the probability of sampling sequence $x$ can be computed
\begin{equation}
	p(x|\theta)=\prod_{n=1}^{N} p(x_n|k(x(n-1)),\theta)=\frac{\prod_{i=1}^W a_i^{(\delta,k_i)}}{A_0^{(\delta,N)}}\,,
\end{equation}
where the function $m^{(\delta,r)}$ is defined as
\begin{equation}
	m^{(\delta,r)}\equiv m(m+\delta)(m+2\delta)\cdots(m+(r-1)\delta)\,.
\end{equation}
Note that $m^{(\delta,r)}$ generalises the multinomial law,
\begin{equation}
	\left(\sum_i a_i\right)^{(\delta,N)}=\sum_{\{k|N=\sum_i k_i\}}{N\choose{k}}\prod_{i=1}^W a_i^{(\delta,k_i)}\,,
\label{multilaw}
\end{equation}
and forms a one-parameter generalisation of powers $m^r$. 
For $\delta=0$,  $m^{(0,r)}=m^r$ and for $\delta=1$, $m^{(1,r)}=(m+r-1)!/(m-1)!$.

The probability of observing a particular histogram $k$ after $N$ trials becomes
\begin{equation}
P(k|\theta,N)={N\choose{k}}\frac{ \prod_{i=1}^W a_i^{(\delta,k_i)} }{ A_0^{\left(\delta,N\right)}}\, ,
\label{PolyaP}
\end{equation}
with
$\sum_{\{k_{i}\geq 0 | \sum_i k_i=N\}}P(k|\theta,N) =1$.
Note that $P(k|\theta,N)$ is almost of multinomial form, it is a multinomial factor times a term depending on $\theta$.
One might conclude that the max-ent functional for P\'olya processes is Shannon entropy 
in combination with a generalised cross-entropy term that depends on $\theta$.
However, this turns out to be wrong, since contributions from the generalised powers $m^{(\delta,r)}$ 
in \equa (\ref{PolyaP}) cancel the multinomial factor almost completely. To see this we first rewrite 
\begin{equation}
\begin{array}{lcl}
a_i^{(\delta,k_i)}&=&a_i(a_i+\delta)\cdots(a_i+(k_i-1)\delta)\\
&=&(a_i+\delta)\cdots(a_i+k_i\delta)\frac{a_i}{a_i+k_i\delta}\\
&=&k_i!\delta^{k_i} (1+\frac{a_i}{\delta})\cdots(1+\frac{a_i}{k_i\delta})\frac{a_i}{a_i+k_i\delta} \\
%&=&\delta^{k_i-1}(k_i-1)!a_i(1+\frac{a_i}{\delta})\cdots(1+\frac{a_i}{(k_i-1)\delta})\\
%&\sim&\delta^{k_i-1}(k_i-1)!\frac{(k_i+1)^{\frac{a_i}{\delta}}}{1+\frac{a_i}{k_i \delta}}\,,
%&\sim&\delta^{k_i-1}(k_i-1)!a_ik_i^{\frac{a_i}{\delta}}\,,
&\sim&k_i!\delta^{k_i}(k_i+1)^{\frac{a_i}{\delta}}\frac{a_i}{a_i+k_i\delta}\,,
\end{array}
\label{approx}
\end{equation}
where we use $\sum_{r=1}^{s}\frac1r\sim\log(s+1)$ and $1+y\sim \exp(y)$, which is valid for sufficiently 
small $y=a_i/\delta$, i.e. for sufficiently large $\delta$. 
With the notation $\gamma\equiv\delta/A_0$ we obtain  
\begin{equation}
%P(k|\theta,N)=\frac{N^{-W-1}}{\left(\gamma+\frac1N\right) \left(1+\frac1N\right)^{\frac1\gamma}}\prod_{i=1}^W \frac{\left(p_i+\frac1N\right)^{\frac{q_i}{\gamma}}}{\frac{p_i}{q_i}\gamma+\frac1N}\,. 
P(k|\theta,N)=\frac{1+\gamma N}{\left(1+\frac1N\right)^{\frac1\gamma}} 
\prod_{i=1}^W \frac{\left(p_i+\frac1N\right)^{\frac{q_i}{\gamma}}}{1+N\gamma\frac{p_i}{q_i}}\,, 
\end{equation}
%
% RUDI in
where $k=pN$.
% RUDI out
Following the construction discussed above, we identify $\Psi(p|\theta,N)=\log P(pN|\theta,N)$, 
% RUDI in
%($k=Np$), 
% RUDI out
which no longer scales explicitly with $N$, but $\phi(N)=1$ ($c=0$), so that $\psi=\Psi$.
% RUDI in
Inserting Eq. (\ref{approx}) into Eq. (\ref{PolyaP}), leads to the expression
\begin{equation}
\begin{array}{lcl}
	\psi(p|\theta)&=&-\sum_{i=1}^W\left(1-\frac{q_i}{\gamma}\right)\log\left(p_i+\frac1N\right)\\
&&-\sum_{i=1}^W\log\left(1+\frac{1}{N\gamma}\frac{q_i-\gamma}{p_i+\frac1N}\right)+\sum_{i=1}^W\log q_i\\
&&-\frac1\gamma\log\left(1+\frac1N\right)+\log\left(1+N\gamma\right)-W\log(N\gamma)\,.
\end{array}
	\label{Eq:GenDiv0}
\end{equation}
%However, in the limit $N\to\infty$
%we obtain the asymptotic result for the generalised "P\'olya information divergence" for large $N$,
More precisely, the {\em finite size} P\'olya ``entropy'' can be conveniently identified with the terms
in $\psi(p|\theta)$ that do not depend on $q$, 
\begin{equation}
% RUDI in
	H^{\rm P\acute{o}lya}(p)=-\sum_{i=1}^W\log\left(p_i+\frac{\lambda}{N}\right)\,,
%	H^{\rm P\acute{o}lya}(p)=-\sum_{i=1}^W\log\left(p_i+\frac{1}{NW\gamma}\right)\,.
%	H^{\rm P\acute{o}lya}(p)=-\sum_{i=1}^W\log\left(p_i+\frac1N\right)\,.
% RUDI out
\label{polyaent2}
\end{equation}
where $\lambda>0$ can in principle be chosen freely. 
Up to a constant depending only on $\gamma$ and $N$, the finite size cross-entropy can be identified with
\begin{equation}
% RUDI in
%		H_{\rm cross}^{\rm P\acute{o}lya}(p|q)=-\sum_i \left[\frac1\gamma q_i\log \left(p_i+1/N\right)
%		-\log\left(1-\frac{1-Wq_i}{1+\gamma WNp_i}\right)\right]\ .
		H_{\rm cross}^{\rm P\acute{o}lya}(p|q)=-\sum_i \left[\frac{q_i}\gamma \log \left(p_i+\frac1N\right)
		-\log\left(1+\frac1{N\gamma}\frac{q_i-\lambda\gamma}{p_i+\frac\lambda{N}}\right)+\log q_i\right]\ .
% RUDI out
\end{equation}
Convenient choices for $\lambda$ are the following. $\lambda=1$, represent $\log \psi$ 
as in Eq. (\ref{Eq:GenDiv0}). Alternatively, one may choose $\lambda=1/(W\gamma)$, which is a convenient choice 
if one considers a uniform initial distribution, $q_i=1/W$, of balls in the urn.
The finite size P\'olya entropy \equa (\ref{polyaent2}), 
yields well defined entropy even if some states $i$ have
vanishing probability $p_i=0$. 

To simplify the following analysis we consider the limit $N\to\infty$ of this functional, 
where the notion of "information divergence" for P\'olya processes, 
essentially reduces to 
% RUDI out
\begin{equation}
	\psi(p|\theta)=-\sum_{i=1}^W\log p_i+\frac1\gamma\sum_{i=1}^W q_i\log p_i+\sum_{i=1}^W\log q_i\,,
	\label{Eq:GenDiv}
\end{equation}
% RUDI in
up to terms of order $1/N$ and terms that do not explicitly depend on $p_i$ or $q_i$.
% RUDI out
%
In this limit the asymptotic P\'olya ``entropy'' is given by, 
\begin{equation}
	H^{\rm P\acute{o}lya}(p)=-\sum_{i=1}^W\log p_i\,. 
\label{polyaent}
\end{equation}
We observe that one cannot derive $H^{\rm P\acute{o}lya}(p)$ from the multiplicity of the system, which gets canceled by counter terms, as we have seen above.
In addition, we note that the $q$ dependent terms, $\sum_i q_i\log p_i$, in equation (\ref{Eq:GenDiv}) play the role of the P\'olya "cross-entropy", which is no longer linear in $p$.

Maximising $\psi(p|\theta)$ with respect to $p$ on $\sum p_i=1$, either leads  to the solution
\begin{equation}
	p_i=\frac1\zeta\left(q_i-\gamma\right)\, , 
\label{interiorsolution}
\end{equation}
for $0< p_i< 1$, or, if this can not be satisfied, to boundary solutions $p_i=0$.  
$\zeta$ is a normalisation constant.
There exist three scenarios: \\

\noindent (A) For $\gamma<\min(q)$, \equa (\ref{interiorsolution}) is the 
max-ent solution for all $i$ (no boundary-solutions). The limit 
$\gamma\to 0$ provides the correct multinomial limit $p_i\to q_i$.

\vspace{1mm}
\noindent (B) If $\max(q)>\gamma>\min(q)$, $\psi$ gets maximal for those $i$ with 
$q_i>\gamma$ and follows solution \equa (\ref{interiorsolution}); those 
$i$ where $q_i<\gamma$ are boundary-solutions, $p_i=0$.

\vspace{1mm}
\noindent (C) For $\gamma>\max(q)$ all $p_i$ are boundary-solutions, meaning that 
one winner $i$ takes it all, $p_i=1$, while all other states have vanishing probability.\\

\begin{figure}[t]
	\centering
		\includegraphics[width=0.8\columnwidth]{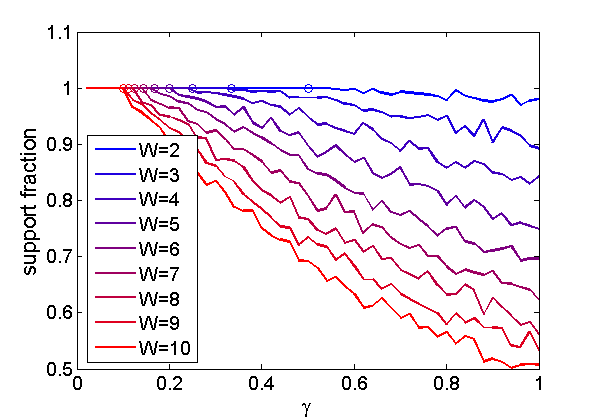}
	\caption{The fraction of distinct colors contained in the P{\'o}lya urn, which at least get sampled once within the first $N=500$ steps
	of the process, for numbers of colors $W=2,3,\cdots,10$ for uniformly distributed initial conditions $q_i=1/W$, $i=1,\cdots,W$, evaluated 
	from $250$ runs for each $\gamma=0.01, 0.02,\cdots,1$. 
	The onset of instability $\gamma_{\rm crit}(q)=\min(q)=1/W$ (circular markers) is very well reproduced experimentally.   
	\label{fig:instab}
	}
\end{figure}

Since $\partial^2 \psi_{\rm P\acute{o}lya}/\partial p_i^2<0\ $if $\gamma < q_i$,  $\psi_{\rm P\acute{o}lya}$ is maximal and case (A) applies. 
If $q_i<\gamma$, Eq (\ref{interiorsolution}) becomes negative but also unstable and is replaced by a boundary solution: cases (B) and (C).
The P\'olya max-ent not only allows us to predict $p_i$ from the initial prior probabilities $q_i$,  
it also identifies $\gamma$ as the crucial parameter that distinguishes between the three regimes of P\'olya urn dynamics\footnote{
% RUDI in
Note that a P{\'o}lya urn $U_1$ that initially contains $A_0$ balls and has evolved for $N$ steps with $\gamma=\delta/A_0$, can be regarded as
another P{\'o}lya urn, $U_2$, in its initial state, containing $A_N=A_0+\delta N$ balls, that evolves with an effective reinforcement parameter 
$\gamma(N)=\delta/A_N$, and the initial distribution of balls $q(N)=p(i|k(N),\theta)$, where $k(N)$ is the histogram of colors drawn in the first $N$ steps
of the original urn process $U_1$. Obviously the asymptotic behavior of P{\'o}lya urns gets determined early on in the process, where the effective reinforcement parameter $\gamma(N)$ is largest. The probability of a P{\'o}lya urn to enter a winner-takes-all dynamics, 
i.e. to end up in one of the scenarios A, B, or C, depends on the reinforcement parameter $\gamma$.}.
% RUDI out
%
For sufficiently large but finite $N$, the analysis above is more involved but solvable.

Assuming uniformly distributed priors, $q_i=1/W$ for all $i$, the max-ent result \equa (\ref{interiorsolution})
correctly predicts uniformly distributed $\hat p_i=1/W$, while observed distributions 
$p$ may strongly deviate from this prediction. This result reflects the fact that despite the P{\'o}lya urn process
being inherently instable (e.g. winner takes all) with little chance of predicting who in particular will win, 
i.e. which color of balls will dominate the others, repeating the experiment many times every color of balls has 
the same chance to win (or biased according to the priors $q$). This discrepancy between ensemble average and 
time average makes it impossible to predict who in particular will win or loose in the course of time.
However, using detailed information about the process one can predict {\em how} winners win. 
In particular one can 
% Rudi in
(i) predict the onset of instability, i.e. the emergence of colors $i$ that will effectively never be drawn, 
at $\gamma_{\rm crit}=\min(q)$ (compare Fig. (\ref{fig:instab})), and 
(ii) construct a maximum entropy functional for predicting the time dependent
frequency distribution of a process, i.e. the number of times one observes states $i$ for $n$ times.
As a consequence, one also can derive the rank distributions of the process, i.e. the frequency 
of observing balls of some color after ranking those frequencies according to their magnitude.

%%%%%%%%%%%%%%%%%%%%%%%%%%%%%%%%%%%%%%%%
\subsection{Rank and frequency distributions of P\'olya urns}
%%%%%%%%%%%%%%%%%%%%%%%%%%%%%%%%%%%%%%%%

With the presented max-ent approach we now compute frequency distribution functions. 
Given the histogram $k = (k_1,k_2, \cdots k_W)$ is obtained after $N$ iterations of the process,we define new variables, 
\begin{equation}
	n_z(k)=\sum_{i=1}^W \chi(k_i=z)\,,
\label{newvar}
\end{equation}
where $\chi$ is the characteristic function that returns $1$ if the argument is true and $0$ if false.
$n_z(k)$ is the number of colors $i$ that occur $z$ times after running the P\'olya  process for $N$ iterations.
$n_z$ is subject to the two constraints, 
\begin{equation}
	W=\sum_{z=0}^N n_z(k)\quad{\rm and}\quad N=\sum_{z=0}^N n_z(k)z\,,
\label{frqconstr}
\end{equation}
which can be included in the maximization procedure introducing Lagrange multipliers, $\alpha$ and $\beta$.
The probability of observing some $n=(n_1,\cdots,n_N)$ is 
\begin{equation}
	\tilde P(n|\theta,N)=\sum_{\{k_i\geq 0|n=n(k)\}}P(k|\theta,N)\,. 
\label{Pk2Pn}
\end{equation}
Defining the relative frequencies $\pi_z=n_z/W$ and $\bar p_z=z/N$
we can construct the max-ent functional from $\tilde P(n|\theta, N)$.
We identify $\tilde \psi(\pi|\theta,N)\equiv \log(\tilde P(n|\theta,N))/W$. 

For the multinomial $P(k|\theta,N)={N\choose{k}}\prod_i q_i^{k_i}$, and uniform  priors $q_i=1/W$ we find 
up to an additive constant,
\begin{equation}
\begin{array}{lcl}
\tilde\psi(\pi|\theta,N)
%&\equiv&\frac1W\log \tilde P(n|\theta)\\ && \\
%&=&\sum_{\{k|n=n(k)\}}P(k|1/W)\\
&=&-\sum_{z=0}^N\pi_z\log \pi_z\\ && \\
&&-N\sum_{z=0}^N\pi_z \bar p_z\log \bar p_z\,.
%-\frac{N}{W}\log W\,.
\end{array}
\end{equation}
$\tilde\psi(\pi|\theta,N)$ has to be maximised subject to \equa (\ref{frqconstr}),
\begin{equation}
\sum_{z=0}^N \pi_z=1\quad{\rm and}\quad \sum_{z=0}^N \pi_z \bar p_z=\frac1W\,,
\label{constraints}
\end{equation}
so that we get the asymptotic solution for large $W$ and large $N$, ($N\gg W\gg 1$), 
\begin{equation}
	\pi_z=\frac{ \phi^z  }{\zeta\ z!}\, . %\quad \phi=\left( T e^{-(1+\frac{\beta}{T})\,,
\label{multinomsol}
\end{equation}
This is the Poisson distribution, exactly as expected for multinomial processes. 
$\phi=N e^{-(1+\frac{\beta}{N})}$, $\zeta$ is a normalisation constant, 
and $\pi_z$ gets maximal at $\hat z=\phi\sim N/W$.

For the P\'olya urn with uniform priors we get from \equa (\ref{Pk2Pn})
\begin{equation}
	\tilde P(n|\theta,N) = \frac1{Z(\theta,N)}\frac{W!}{\prod_{z=0}^N n_z!} 
	\prod_{z=0}^N \left[\frac{ \left(\bar p_z+\frac1N\right)^{\frac{1}{W\gamma}} }{ \bar p_z + \frac{1}{\gamma WN} } \right]^{n_z} \, .
\end{equation}
 $Z(\theta,N)$ is the normalisation.
Up to a constant the max-ent functional $\tilde\psi_{\rm P\acute{o}lya}(\pi|\theta,N)\equiv \log(P(n|\theta,N))/ W$ is 
\begin{equation}
\begin{array}{lcl}
\tilde\psi_{\rm P\acute{o}lya}(\pi|\theta,N)&=&-\sum_{z=0}^N \pi_z\log(\pi_z)\\ && \\
&&-\sum_{z=0}^N \pi_z \log\left(\bar p_z + \frac{1}{\gamma WN}\right)\\ && \\
&&+\frac1{W\gamma}\sum_{z=0}^N \pi_z\log\left(\bar p_z+\frac1N\right)\,.
\end{array}
\end{equation}
maximising $\tilde\psi_{\rm P\acute{o}lya}(\pi|\theta,N)$ under the conditions of \equa (\ref{constraints}) 
provides the frequency distribution of the P\'olya process for uniform priors,
\begin{equation}
	\pi_z=\frac1\zeta\phi^z\frac{(z+1)^{\frac1{W\gamma}}}{z+\frac{1}{\gamma W}}\,,
\label{polyasolution}
\end{equation}
with $\phi=\exp(-\beta)$, and normalisation $\zeta=\exp(1+\alpha)N^{\frac1{W\gamma}-1}$. 
%
%The frequency distribution of P\'olya processes does not in general  
%concentrate around $N/W$, for $N\to\infty$, such as multinomial processes do, 
%with typical deviations from the uniform distribution of order $\sqrt{N}$.
%
\begin{figure}[t]
	\centering
		\includegraphics[width=0.8\columnwidth]{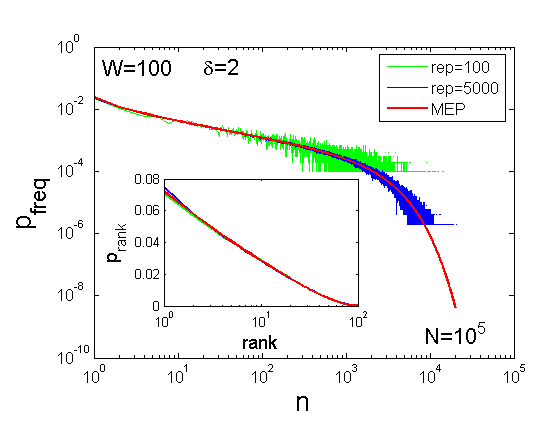}
	\caption{Frequency distribution of a P\'olya urn process and uniform initial conditions (red line), for $W=100$, $\delta=2$, 
	$a_i=1$ for all $i=1$, and $N=10^5$ steps. Simulations are shown for $100$ (green) and $5000$ (blue) repetitions of the process. 
	Inset: Rank distributions of the max-ent result and the numerical realisations in semi-log scale.   
	\label{fig:PolyaUrnMEP}
	}
\end{figure}

The rank distribution of states, $f(r)$, can now be obtained as follows.
$r=1$ is the state that occurs most frequently, $r=W$ is the least occupied state.  
For $r=1,\cdots,W$ we define intervals $[t_{r+1} t_{r}]$ with
$t_1=N$ and $t_{W+1}=0$, such that $\sum_{t_{r+1}\leq z<t_{r}} \pi_z\sim 1/W$. 
To find $t_i$ we substitute sums by integrals and get
\begin{equation}
	\frac1W=\int_{t_{r+1}}^{t_{r}}dz\ \pi_z\quad{\rm and}\quad f(r)=\frac{W}{N}\int_{t_{r+1}}^{t_{r}}dz\ \pi_z z \, . 
\label{ranktransformation}
\end{equation}
Results for the frequency distributions for $a_i=1$, $W=100$, and $\delta=2$ are shown in Fig. (\ref{fig:PolyaUrnMEP}), together with a numerical 
simulation for the same process. The inset shows the rank distribution. The P\'olya max-ent predicts frequency and rank distribution extremely well. 

The above results were all derived under the assumption that $\gamma>0$ is sufficiently large.
By numerical simulation we find that the solution \equa (\ref{polyasolution}) also works remarkably well 
for very small values of $\gamma$, if the value of $\gamma$ in \equa (\ref{polyasolution})
is appropriately renormalised, $\gamma \to \gamma_0$. In particular for $\gamma=0$ (multinomial process)
we sample the Poisson distribution function, \equa (\ref{multinomsol}). The P\'olya max-ent solution 
recovers the Poisson distribution extremely well %(up to second order around the maximum)
if  $\gamma=0 \to \gamma_0(W,N)= 1/(N+3W)$. 
In this sense the P\'olya max-ent remains adequate in the limit of small $\gamma$.

\section{Discussion}

P\'olya urns offer a transparent way to study self-reinforcing systems with explicit path-dependence. 
Based on the microscopic rules of the process, we constructively derive the generalised 
information divergence $\psi$ which acts as the corresponding non-multinomial max-ent functional. This 
provides us with an alternative to the {\em ensemble approach} for path-dependent processes that is able to predict 
the statistics of the system. The maximisation of the functional leads to an equivalent to the classical {\em maximum configuration} approach, 
which by definition predicts the most likely distribution function. In this sense maximum configuration predictions are optimal, 
and can be used to understand even details of the statistics of path-dependent processes, such as their frequency and rank distributions.   

It is interesting to note that the functional playing the role of the entropy in the  P\'olya processes violates at least 
two of the four classic information theoretic (Shannon-Khinchin) axioms 
which determine Shannon entropy \cite{Khinchin}. Even more, for the finite size P\'olya entropy, three of the four axioms are violated. 
This indicates that the classes of generalised entropy functionals that are useful for a max-ent approach may 
be even larger than expected \cite{HTclassification,HThowphasespacegrows}. 
One might speculate that in this sense the classic information theoretic axioms are too rigorous, when it comes to 
characterising information flow and phase space structure in non-stationary, path-dependent, processes.  
The observation that each particular class of non-multinomial processes requires a matching max-ent functional 
that can in principle be constructed from the generative rules of a process, opens the applicability of max-ent approaches for a wide
range of complex systems in a meaningful way. The generalized max-ent approach in this sense responds
to Einsteins comment on Boltzmann's principle in a natural way.
%It might even trigger a discussion about the possibility of an information theory for 
%complex processes.

Finally we note the implications for statistical inference with data from non-multinomial sources, 
which implicitly involves the estimation of the parameters $\theta$ that determine the process that generates the data. 
In a max-ent approach this is done by fitting classes of curves to the data, that are consistent with the max-ent approach. 
For doing this, the nature of the process, i.e. its class, needs to be known. 
For path-dependent processes, which are non-multinomial by nature, entropy will no longer be Shannon entropy $H$, 
and the information divergence will no longer be the Kullback-Leibler divergence. 
\newline


\begin{thebibliography}{10}
%%%%%%%%%%%%%%%%%%%%%%%%%%%%%%%%%%%%%%%%%%%%%%%%%%%%%%%%  
%%%%%%%%%%%%%%%%%%%%%%%%%%%%%%%%%%%%%%%%%%%%%%%%%%%%%%%%  
\expandafter\ifx\csname url\endcsname\relax
  \def\url#1{\texttt{#1}}\fi
\expandafter\ifx\csname urlprefix\endcsname\relax\def\urlprefix{URL }\fi
\providecommand{\bibinfo}[2]{#2}
\providecommand{\eprint}[2][]{\url{#2}}
%%%%%%%%%%%%%%%%%%%%%%%%%%%%%%%%%%%%%%%%%%%%%%%%%%%%%%%%  
%%%%%%%%%%%%%%%%%%%%%%%%%%%%%%%%%%%%%%%%%%%%%%%%%%%%%%%%  

\bibitem{Einstein:1910}
Einstein A 1910, Theorie der Opaleszenz von homogenen Fl{\"u}ssigkeiten und 
Fl{\"u}ssigkeitsgemischen in der N{\"a}he des kritischen Zustandes. 
{\em Ann. d. Phys. {\bf 33}, 1275}.


\bibitem{LandauLifshitz}
Landau L D and Lifshitz E M 1980,
{\em Statistical Physics, Course in Theoretical Physics {\bf 5}}, 
footnote p. 12. (Elsevier, 3'rd edition)


\bibitem{OPeters}
Peters O. 2011, 
The time resolution of the St Petersburg paradox.
{\em Phil. Trans. R. Soc. A 396, 4913-4931}
%and O. Peters and M. Gell-Mann, Chaos 26, 023103 (2016)). 

%15
\bibitem{HTMGM3}
Hanel R, Thurner S, and Gell-Mann M 2014, 
How multiplicity of random processes determines entropy: derivation of the maximum entropy principle for complex systems,
{\em Proc. Nat. Acad. Sci. USA {\bf 111} 6905--6910}. 

%13
\bibitem{Shannon1948}
Shannon C E 1948,
A Mathematical Theory of Communication,
{\em Bell Syst. Tech. J. {\bf 27} 379–-423, 623–-656}.

%14
\bibitem{Jaynes1968} 
Jaynes E T 1968, 
Prior Probabilities, 
{\em IEEE Trans Sys Sci and Cybernetics} {\bf 4} 227-–241.

%26
\bibitem{KullbackLeibler}
Kullback S, and Leibler R A 1951,
On information and sufficiency, 
{\em Ann. Math. Stat. {\bf 22} 79-–86}.

%16
\bibitem{Khinchin}
Khinchin A I 1957, 
{\em Mathematical foundations of information theory},
(Dover Publ., New York).

%1
\bibitem{Polya1923}
Eggenberger F, G. P\'olya G 1923, 
{\"U}ber die Statistik verketteter Vorg{\"a}nge,
{\em Z. Angew. Math. Mech. {\bf 1} 279--289}.

%2
\bibitem{Polya:1930}
P\'olya G 1930, 
Sur quelques points de la th\'eorie des probabilit\'es, 
{\em Ann. Inst. Henri Poincare {\bf 1} 117--161}.

%17
\bibitem{Wallstrom2012}
Wallstrom T C 2012,
The equalization probability of P\'olya urn,
{\em Am. Math. Mon. {\bf 119} 516--518}. 

%18
\bibitem{Johnson:1977}
Johnson N L and Kotz S 1977,  
Urn Models and Their Application: An Approach to Modern Discrete Probability Theory,  
In {\em Urn models and their application. An approach to modern discrete probability theory}. (John Wiley, New York).

%19
\bibitem{Mahmoud:2008}
Mahmoud H 2008, 
P\'olya Urn Models, 
{\em Texts in Statistical Science},
(Chapman \& Hall/CRC Texts in Statistical Science, Taylor and Francis Ltd, Hoboken, NJ).

%20
\bibitem{Kotz2000}
Kotz S, Mahmoud H, and Robert P 2000,
On generalised P\'olya urn models, 
{\em Stat. Prob. Lett. {\bf 49} 163--173}. 

%21ß
\bibitem{Janson2004}
Janson S 2004,
Functional limit theorems for multitype branching processes and generalised P\'olya urns,
{\em Stoch. Proc. Appl. {\bf 110} 177–-245}.

%22
\bibitem{Smythe1996}
Smythe R T 1996,
Central limit theorems for urn models,
{\em Stoch. Proc. Appl. {\bf 65} 115--137}.

%23
\bibitem{Gouet1993}
Gouet R 1993,
Martingale Functional Central Limit Theorems for a generalised P\'olya Urn,
{\em Ann. Prob. {\bf 21} 1624--1639}.

%3
\bibitem{Tolusso2011}
Tolusso D and Wang X 2011,
Interval estimation for response adaptive clinical trials,
{\em Comput. Stat. Data Anal. {\bf 55} 725--730}.

%4
\bibitem{Binder2009}
Binder B J and Landman K A 2009,
Tissue growth and the P\'olya distribution,
{\em Aust. J. Eng. Edu. {\bf 15} 35--42}. 
   
%5    
\bibitem{Crouch2004}
Crouch C and Farrell H 2004,    
Breaking the Path of Institutional Development? Alternatives to the New Determinism,
{\em Ratio. Soc. {\bf 16} 5--43}.    

%6    
\bibitem{Bagchi1985}
Bagchi A and Pal A K 1985,  
Asymptotic Normality in the generalised P\'olya–Eggenberger Urn Model, with an Application to Computer Data Structures,
{\em SIAM J. Algeb. Disc. Meth. {\bf 6} 394–-405}.

%7
\bibitem{Geppert2012}
Geppert T 2012, 
EU-Agrar- und Regionalpolitik, Wie vergangene Entscheidungen zuk{\"u}nftige Entwicklungen
beeinflussen - Pfadabh{\"a}ngigkeit und die Reformf{\"a}higkeit von Politikfeldern,
{\em PhD Thesis}, (University of Bamberg Press, Bamberg). 
%http://www.uni-bamberg.de/ubp/

%8 
\bibitem{Donnelly1986}
Donnelly P 1986,
Partition structures, P\'olya urns, the Ewens sampling formula, and the ages of alleles, 
{\em Theor. Popul. Biol. {\bf 30} 271-–288}.

%9
\bibitem{Hoppe1984}
Hoppe F M 1984,
P\'olya-like urns and the Ewens' sampling formula,
{\em J. Math. Biol. {\bf 20} 91--94}.

%10
\bibitem{Banerjee1999}
Banerjee A, Burlina P, and Alajaji F 1999, 
Image segmentation and labeling using the P\'olya urn model, 
{\em IEEE Trans. Image. Proc. {\bf 8} 1243--1253}.  

%11  
\bibitem{Alexander:2012}
Alexander J M, Skyrms B, and Zabell S 2012, 
Inventing new signals, 
{\em Dyn. Games Appl. {\bf 2} 129--145}.

%12
\bibitem{Tria:2014}
Tria F, Loreto V,Servedio	V D P, and Strogatz S H 2014,  
The dynamics of correlated novelties,
{\em Sci. Rep. {\bf 4}:5890}.

%24  
\bibitem{Grendar2010}
Grendar M, and Niven R K 2010,
The P\'olya information divergence,
{\em Info. Sci. {\bf 180} 4189-–4194}. 
  
%25  
\bibitem{Sanov1957}
Sanov I N 1957,  
On the probability of large deviations of random variables, 
{\em Mat. Sbornik {\bf 42} 11--44}.  

%27
\bibitem{HTclassification}
Hanel R and Thurner S 2011, 
A comprehensive classification of complex statistical systems and an ab initio derivation of their entropy and distribution function,
{\em EPL {\bf 93}:20006}.

%28
\bibitem{HThowphasespacegrows}
Hanel R and Thurner S 2011,  
When do generalised entropies apply? How phase space volume determines entropy,
{\em EPL {\bf 96}:50003}.

%%%%%%%%%%%%%%%%%%%%%%%%%%%%%%%%%%%%%%%%%%%%%%%%%

%%1
%\bibitem{Polya1923}
%Eggenberger F, G. P\'olya G (1923), 
%{\"U}ber die Statistik verketteter Vorg{\"a}nge,
%{\em Z. Angew. Math. Mech. {\bf 1} 279--289}.
%
%%2
%\bibitem{Polya:1930}
%P\'olya G (1930), 
%Sur quelques points de la th\'eorie des probabilit\'es, 
%{\em Ann. Inst. Henri Poincare {\bf 1} 117--161}.
%
%%3
%\bibitem{Tolusso2011}
%Tolusso D and Wang X (2011),
%Interval estimation for response adaptive clinical trials,
%{\em Comput. Stat. Data Anal. {\bf 55} 725--730}.
%
%%4
%\bibitem{Binder2009}
%Binder B J and Landman K A (2009),
%Tissue growth and the P\'olya distribution,
%{\em Aust. J. Eng. Edu. {\bf 15} 35--42}. 
%   
%%5    
%\bibitem{Crouch2004}
%Crouch C and Farrell H (2004),    
%Breaking the Path of Institutional Development? Alternatives to the New Determinism,
%{\em Ratio. Soc. {\bf 16} 5--43}.    
%
%%6    
%\bibitem{Bagchi1985}
%Bagchi A and Pal A K (1985),  
%Asymptotic Normality in the generalised P\'olya–Eggenberger Urn Model, with an Application to Computer Data Structures,
%{\em SIAM J. Algeb. Disc. Meth. {\bf 6} 394–-405}.
%
%%7
%\bibitem{Geppert2012}
%Geppert T (2012), 
%EU-Agrar- und Regionalpolitik, Wie vergangene Entscheidungen zuk{\"u}nftige Entwicklungen
%beeinflussen - Pfadabh{\"a}ngigkeit und die Reformf{\"a}higkeit von Politikfeldern,
%{\em PhD Thesis}, (University of Bamberg Press, Bamberg). 
%%http://www.uni-bamberg.de/ubp/
%
%%8 
%\bibitem{Donnelly1986}
%Donnelly P (1986),
%Partition structures, P\'olya urns, the Ewens sampling formula, and the ages of alleles, 
%{\em Theor. Popul. Biol. {\bf 30} 271-–288}.
%
%%9
%\bibitem{Hoppe1984}
%Hoppe F M (1984),
%P\'olya-like urns and the Ewens' sampling formula,
%{\em J. Math. Biol. {\bf 20} 91--94}.
%
%%10
%\bibitem{Banerjee1999}
%Banerjee A, Burlina P, and Alajaji F (1999), 
%Image segmentation and labeling using the P\'olya urn model, 
%{\em IEEE Trans. Image. Proc. {\bf 8} 1243--1253}.  
%
%%11  
%\bibitem{Alexander:2012}
%Alexander J M, Skyrms B, and Zabell S (2012), 
%Inventing new signals, 
%{\em Dyn. Games Appl. {\bf 2} 129--145}.
%
%%12
%\bibitem{Tria:2014}
%Tria F, Loreto V,Servedio	V D P, and Strogatz S H (2014),  
%The dynamics of correlated novelties,
%{\em Sci. Rep. {\bf 4}:5890}.
%
%%13
%\bibitem{Shannon1948}
%Shannon C E (1948),
%A Mathematical Theory of Communication,
%{\em Bell Syst. Tech. J. {\bf 27} 379–-423, 623–-656}.
%
%%14
%\bibitem{Jaynes1968} 
%Jaynes E T (1968), 
%Prior Probabilities, 
%{\em IEEE Trans Sys Sci and Cybernetics} {\bf 4} 227-–241
%
%%15
%\bibitem{HTMGM3}
%Hanel R, Thurner S, and Gell-Mann M (2014), 
%How multiplicity of random processes determines entropy: derivation of the maximum entropy principle for complex systems,
%{\em Proc. Nat. Acad. Sci. USA {\bf 111} 6905--6910}. 
%
%%16
%\bibitem{Khinchin}
%Khinchin A I (1957), 
%{\em Mathematical foundations of information theory},
%(Dover Publ., New York).
%
%
%
%%17
%\bibitem{Wallstrom2012}
%Wallstrom T C (2012),
%The equalization probability of P\'olya urn,
%{\em Am. Math. Mon. {\bf 119} 516--518}. 
%
%%18
%\bibitem{Johnson:1977}
%Johnson N L and Kotz S (1977),  
%Urn Models and Their Application: An Approach to Modern Discrete Probability Theory,  
%In {\em Urn models and their application. An approach to modern discrete probability theory}. (John Wiley, New York).
%
%%19
%\bibitem{Mahmoud:2008}
%Mahmoud H (2008), 
%P\'olya Urn Models, 
%{\em Texts in Statistical Science},
%(Chapman \& Hall/CRC Texts in Statistical Science, Taylor and Francis Ltd, Hoboken, NJ).
%
%%20
%\bibitem{Kotz2000}
%Kotz S, Mahmoud H, and Robert P (2000),
%On generalised P\'olya urn models, 
%{\em Stat. Prob. Lett. {\bf 49} 163--173}. 
%
%%21
%\bibitem{Janson2004}
%Janson S (2004),
%Functional limit theorems for multitype branching processes and generalised P\'olya urns,
%{\em Stoch. Proc. Appl. {\bf 110} 177–-245}.
%
%%22
%\bibitem{Smythe1996}
%Smythe R T (1996),
%Central limit theorems for urn models,
%{\em Stoch. Proc. Appl. {\bf 65} 115--137}.
%
%%23
%\bibitem{Gouet1993}
%Gouet R (1993),
%Martingale Functional Central Limit Theorems for a generalised P\'olya Urn,
%{\em Ann. Prob. {\bf 21} 1624--1639}.
%
%%24  
%\bibitem{Grendar2010}
%Grendar M, and Niven R K (2010),
%The P\'olya information divergence,
%{\em Info. Sci. {\bf 180} 4189-–4194}. 
%  
%%25  
%\bibitem{Sanov1957}
%Sanov I N (1957),  
%On the probability of large deviations of random variables, 
%{\em Mat. Sbornik {\bf 42} 11--44}.  
%
%%26
%\bibitem{KullbackLeibler}
%Kullback S, and Leibler R A (1951),
%On information and sufficiency, 
%{\em Ann. Math. Stat. {\bf 22} 79-–86}.
%
%%27
%\bibitem{HTclassification}
%Hanel R and Thurner S (2011), 
%A comprehensive classification of complex statistical systems and an ab initio derivation of their entropy and distribution function,
%{\em EPL {\bf 93}:20006}.
%
%%28
%\bibitem{HThowphasespacegrows}
%Hanel R and Thurner S (2011),  
%When do generalised entropies apply? How phase space volume determines entropy,
%{\em EPL {\bf 96}:50003}.




%%%%%%%%%%%%%%%%%%%%%%%%%%%%%%%%%%%%%%%%%%%%%%%%%%%%%%%%  
\end{thebibliography}
\end{document}